\documentclass[10pt,journal]{IEEEtran}

\usepackage{framed}
\usepackage{cite} 
\usepackage[table]{xcolor}
\usepackage{graphicx}
\usepackage{booktabs}
\usepackage{caption}
\usepackage{subcaption}
\usepackage{amssymb}
\usepackage{comment}
\usepackage{gensymb}
\usepackage{enumitem}
\usepackage{colortbl}
\usepackage{multirow}
\usepackage{hyperref}
\hypersetup{
    colorlinks,
    linkcolor={red!50!black},
    citecolor={blue!50!black},
    urlcolor={blue!80!black}
}

\usepackage{pifont}% http://ctan.org/pkg/pifont
\definecolor{brightpurple}{RGB}{150, 61, 255}

\begin{document}

\title{Training Spatial Ability in Virtual Reality}
\author{\emph{Yiannos Demetriou, Manasvi Parikh, Sara Eskandari, Westley Weimer, Madeline Endres}}

\maketitle

% FIXME - remove bolding
\begin{abstract}
\noindent \textbf{Background:} Spatial reasoning has been identified as a critical skill for success in STEM. Unfortunately, 
under-represented groups often have lower incoming spatial ability. Courses that improve spatial skills exist but are not widely used. Virtual reality (VR) has been suggested as a possible tool for teaching spatial reasoning since students are more accurate and complete spatial tasks more quickly in three dimensions. However, no prior work has developed or evaluated a fully-structured VR spatial skills course.

\vspace{5pt}
\noindent \textbf{Objectives:} We seek to assess the effectiveness of teaching spatial reasoning in VR, both in isolation as a structured training curriculum and also in comparison to traditional methods.

\vspace{5pt}
\noindent \textbf{Methods:} We adapted three modules of an existing pencil-and-paper course to VR, leveraging educational scaffolding and real-time feedback in the design. We evaluated our three-week course in a study with $n=24$ undergraduate introductory STEM students, capturing both quantitative spatial ability gains (using pre- and post test scores on validated assessments) and qualitative insights (from a post-study questionnaire). We also compared our VR course to an offering of a baseline non-VR course (using data collected in a previous study).

\vspace{5pt}
\noindent \textbf{Results and Conclusions:} Students who took our VR course had significant spatial ability gains. Critically, we find no significant
difference in outcomes between our VR course (3 meetings of 120 minutes each) and a baseline pencil and paper course (10 meetings of 90 minutes each), 
suggesting that spatial reasoning can be very efficiently taught in VR. We observed cybersickness at lower rates than are generally reported and most students reported enjoying learning in VR.

\end{abstract}

\iffalse


\fi

\pagestyle{plain} 

\section{Lay Summary}
\label{sec:summary}

\noindent What is currently known about this topic?
\begin{itemize}
 \item Spatial skills can be taught and correlate with improved STEM and CS outcomes.
 \item Pencil-and-paper spatial skills training is effective.
\end{itemize}
What does this paper add?
\begin{itemize}
  \item We design and implement a Virtual Reality spatial skills training. 
  \item It is as effective as pencil-and-paper training.
  \item It is very efficient compared to pencil-and-paper training. 
\end{itemize}
Implications for practice/or policy
\begin{itemize}
  \item Some schools have mandatory spatial skills courses.
  \item Our VR course may be easier to deploy remotely.
  \item Our VR course takes less time to obtain the same benefits.
\end{itemize}

\section{Introduction}
\label{sec:introduction}
Spatial reasoning ability, defined as the capacity to understand and reason about spatial relations, has been found to be a major predictor of student success in STEM (science, technology, engineering, math)~\cite{Margulieux2019, Jones2008, Sorby2009, Wai2009}. Students from groups under-represented in STEM, such as women and students from a lower socioeconomic backgrounds, often have lower incoming spatial ability~\cite{Miranda2018, Segil2016}. For example, differences in incoming spatial ability better explains the under-performance of lower socioeconomic status computing students than does technology access~\cite{Miranda2018}, and higher spatial ability can provide a stronger advantage than confidence or experience~\cite{Jones2008}.

Intriguingly, spatial ability is not fixed, and can be improved through training~\cite{Sorby2009, Sorby2018, Uttal2012}.  Existing courses are effective in increasing spatial ability~\cite{Sorby2009, Sorby2018}. However, spatial ability courses are not yet ubiquitous and can be time-intensive~\cite{Sorby2009}. As a result, many students do not have the opportunity to increase their spatial ability, leading to lower grades, lower diversity, and lower retention rates in STEM~\cite{Sorby2018, Sorby2005, Miranda2018, Segil2016}. In addition, existing courses primarily use traditional instructional media such as paper workbooks or two-dimensional (2D) software~\cite{Sorby2009}. However, this may make teaching spatial reasoning challenging because of the inherent limitations of teaching a three-dimensional (3D) subject using 2D tools. This dimensionality is important: For example, spatial tasks can be completed more accurately and rapidly with 3D stimuli as opposed to 2D stimuli~\cite{lochhead_2022}. To help mitigate this misalignment, some courses provide students physical blocks as a visualization aid~\cite{Sorby2009}, but structuring and giving feedback on their use can be challenging in a classroom setting. 

As a 3D medium, extended reality technologies (e.g., virtual reality (VR) or augmented reality) have the potential to enhance existing spatial reasoning training by more formally and directly supporting 3D reasoning via explicit educational scaffolding and real-time feedback. Recently, there has been excitement regarding the promise of this approach~\cite{Carmona2018, chang2017}. 
Notably, when discussing developments in extended reality technologies, a specialist meeting on spatial thinking in the college curriculum concluded that ``there is a significant need to integrate these developments into the framework of general spatial thinking and skill development''~\cite{Janelle2014}. In addition, it has been found that one's spatial ability can increase simply by using VR for other subjects such as architecture and astrophysics~\cite{DARWISH2023, baumgartner2022, huang2021}. However, to our knowledge, only two studies attempted to specifically teach spatial reasoning in VR so far, and both used short, unstructured sessions~\cite{chang2017, Carmona2018}. A fully-structured VR spatial reasoning course based on, and comparable to, a validated existing traditional course has yet to be developed and tested.
 
In this paper, we work to close this gap through the development and evaluation of an immersive virtual reality tool for training spatial ability. Specifically, we adapt three modules from an established pencil-and-paper course~\cite{Sorby2009} to VR. Beyond making the transition from 2D to 3D more concrete and adapting existing problems to a new medium, we augment the original course with additional educational scaffolding (the structured addition and removal of extra supports to enhance student learning) and interactive real-time feedback. We propose increasing the scaffolding present compared to the baseline course because not only can computer-based scaffolding improve overall learning~\cite{Belland2017}, but also learning  abstract concepts (such as spatial reasoning) are easier when they are scaffolded by concrete visual representations~\cite{Moreno2011}. We also propose including interactive real-time feedback because current research suggests that more elaborate feedback leads to higher learning outcomes~\cite{Kleij2015}, and that disconfirmation without corrective information is of little use~\cite{Hattie2007}. In addition, it has been suggested that feedback is most useful when it helps students understand why they made the mistakes they did and how to avoid them next time~\cite{Wisniewski2020}, a self-reflective process that can be facilitated through interactivity.

We hypothesize that immersive VR technology will be at least as effective as pencil-and-paper and 2D software training because many aspects of spatial reasoning are inherently 3D. We also observe students' engagement levels because studies have found a positive effect on attention and engagement when using instructional immersive VR~\cite{Allcoat2018, Thompson2020, Ruixue2020}. To the best of our knowledge, we are the first to explore teaching spatial reasoning in virtual reality in a structured course environment.

We evaluated the course using $n=24$ university students enrolled in introductory computer science. We found that students in our three-week VR course had significant spatial ability gains ($p=0.007$, $d=0.50$). Of the 24 participants, 20 indicated they enjoyed learning in VR ($83.3\%$). In addition, we compared our VR intervention to a prior offering of the baseline pencil-and-paper course offered to students in a similar introductory computing course. We did not find any statistically-significant differences in outcomes, suggesting that our VR training is as effective as the baseline. However, at 360 minutes (three 120-minute sessions) our VR course is much shorter than the 900-minute pencil-and-paper offering (ten 90-minute sessions). 
 Furthermore, we investigated the impact of our feedback and scaffolding proposals through a qualitative analysis of reported experiences as well as a quantitative analysis of application log data. We found that although higher incoming ability students spent significantly less time on each problem, they did not spend significantly less time looking at feedback. This suggests that our feedback is helpful to both lower and higher incoming ability students alike. With regard to scaffolding, 13 out of 24 students ($54.2\%$) explicitly mentioned finding the 3D visualization enjoyable or helpful in preparing them for 2D problems.

This article's contributions include:

\begin{itemize}
    \item A novel three-week, 360-minute virtual reality spatial reasoning intervention
    \item Quantitative analysis of the learning outcomes after participation in the course
    \item Qualitative insights into how students learn and interact with spatial reasoning in VR
    \item A comparison of the learning outcomes in VR to those in an established physical course
        
\end{itemize}

We conclude with a discussion of the implications of our findings on spatial reasoning education.

\section{Background and Related Work}
\label{sec:background}
We briefly describe the current state of VR in education, how spatial reasoning is related to STEM, and prior work at the intersection of spatial reasoning training and VR.

\subsection{\textbf{Virtual Reality in Education}}
\label{sec:vr_education}
    The term \emph{VR} has been used to represent a very broad set of technologies, including desktop VR, 360-degree videos, augmented reality, and head-mounted displays (HMDs)~\cite{Radianti2020}. Recently, the term \emph{immersive VR}  has been used to describe the VR technologies in which the user cannot distinguish the boundary between the virtual world and the real one (i.e., HMDs)~\cite{Radianti2020}. The use of both non-immersive and immersive VR in education is of great interest, as evidenced by the large number of meta-analyses published since 2020~\cite{Hamilton2020, Radianti2020, Noah2020, Luo2021, Soliman2021, Villena2022}. The idea of using VR in education has been around for decades. Due in part to rapid technological advancements in VR technology, the focus has often been on developing the tools and testing their usability rather than on evaluating educational effectiveness~\cite{Radianti2020, Luo2021}. In a meta-analysis of immersive VR in higher education, Radianti \emph{et al.} found that 46\% of papers focused on user experience alone~\cite{Radianti2020} and not educational outcomes. Similarly, the meta-analyses of Noah \emph{et al.} and Luo \emph{et al.} found that most studies did not measure the impact of VR on learning~\cite{Noah2020, Luo2021}. 
    
    The studies that do assess the impact of VR on learning tend to find a small to medium positive effect for K-12 and higher education~\cite{Luo2021, Jensen2018, Hamilton2020, Villena2022, Soliman2021, Noah2020}. Notably, Hamilton \emph{et al.} found that skills acquired in VR could be successfully transferred to solve real-world problems~\cite{Hamilton2020}. This could explain why a plurality of VR education studies (34\%) used VR for \textit{engineering} education~\cite{Radianti2020}, generally finding both better performance and better education experience with VR as compared to traditional methods~\cite{Soliman2021}. 
    
    %Additionally, a recent study found that a majority use of a new university-wide VR lab was by the Engineering Department\cite{Thomas2022}. 

    % FIMXE: MLE suggests removing for now: Luo \emph{et al.} found that most VR studies that assessed learning outcomes used low levels of immersion (no HMDs) ~\cite{Luo2021}, so the effect of highly immersive VR on learning is still largely unclear despite initial results suggesting larger gains in academic achievement for immersive VR methods as compared to traditional non-immersive or semi-immersive methods~\cite{Villena2022}. %removed this for now: Additionally Hamilton \emph{et al.} found that most studies reported a significant advantage of using IVR, while only two studies reported detrimental effects~\cite{Hamilton2020}. 
    
    Beyond aiding learning, VR has also shown promise in increasing student engagement. For example, Thompson \emph{et al.} added VR to a university-level anatomy and physiology course, finding that students felt more engaged during the VR sessions than during other learning activities \cite{Thompson2020}.
    Similarly, Liu \emph{et al.} used VR to teach science lessons to 6th graders, finding that the VR group had both significantly higher engagement and academic achievement as compared to the traditional method group \cite{Ruixue2020}. They also found that the VR group had high technology acceptance for the use of VR in the classroom. These results suggest that VR can increase the engagement of students of various ages, and that students are accepting of VR methods.
    
% IF NEED ROOM TO CUT, REST OF PARAGRAPH
%They found that over 5 semesters, 4833 students received lessons in the VR lab, and the number of students using the lab increased by 250\% over that time-period. Not only did the instruction through VR quickly gain popularity, but 71\% of the students surveyed also reported enhanced learning outcomes, highlighting the possible increase of both engagement and learning outcomes through the use of VR.

\subsection{\textbf{Spatial Reasoning in STEM}}
\label{sec:spatialAbility_stem}
Spatial reasoning refers to the ability to mentally manipulate three-dimensional objects, and involves skills such as mental rotation and mental folding~\cite{Margulieux2019}. Spatial reasoning ability is impacted by environmental factors~\cite{king2019}. On average, some demographic groups, such as women and students from lower socioeconomic backgrounds, have a lower incoming spatial ability~\cite{Miranda2018, Segil2016, Casey2011, Voyer1995}. 

The role of spatial reasoning in STEM has been studied extensively, establishing significant correlations. In particular, spatial reasoning ability is positively correlated with STEM outcomes~\cite{Wai2009, Alias2002, Fincher2006}. For instance, in a large longitudinal study, Wai \emph{et al.} found that spatial ability ``surfaced as a salient psychological attribute among those adolescents who subsequently go on to achieve advanced educational credentials and occupations in STEM''~\cite{Wai2009}. In another study, Jones \emph{et al.} found that higher spatial ability provides a stronger advantage in early programming than confidence or experience~\cite{Jones2008}. Notably, by using medical imaging, Huang \emph{et al.} found that solving mental rotation problems and tree-based data structure problems in computer science recruited similar brain regions, indicating a deep connection between spatial reasoning and STEM~\cite{Huang2019}. 

Beyond correlational studies, several studies have recently established causality between spatial reasoning training, and improved STEM outcomes~\cite{Sorby2018, Cooper2015, Bockmon2020}. A meta-analysis of spatial training studies found that not only can spatial skills be significantly improved through training, but the training also ``transferred to other spatial tasks that were not directly trained''~\cite{Uttal2012}. For example, in a study with over 3000 first-year engineering students, Sorby \emph{et al.} found that a spatial skills intervention led to improved performance in ``a variety of introductory courses and STEM GPAs, with a particularly positive effect in Engineering Problem Solving and Analysis''~\cite{Sorby2018}. They also found that the intervention positively impacted the retention of women in engineering. Additionally, in a replication of a previous study~\cite{Cooper2015} but using a much larger sample ($n = 345$), Bockmon \emph{et al.} found improved performance on both a spatial skills test and a computer science test, following a spatial skills intervention~\cite{Bockmon2020}.  Thus, since spatial reasoning ability is both malleable and relevant to STEM outcomes, it is important to explore spatial skills interventions as a means to improve STEM outcomes.

Based on this evidence, teaching spatial reasoning to students could have profound effects on their achievements. Therefore, as concluded in a specialist meeting at the University of California, Santa Barbara~\cite{Janelle2014}, it is important that we find new and improved ways to teach spatial reasoning and make spatial skills interventions more appealing to students.

 \begin{figure*}[t]
    \centering
    \includegraphics[width=18cm]{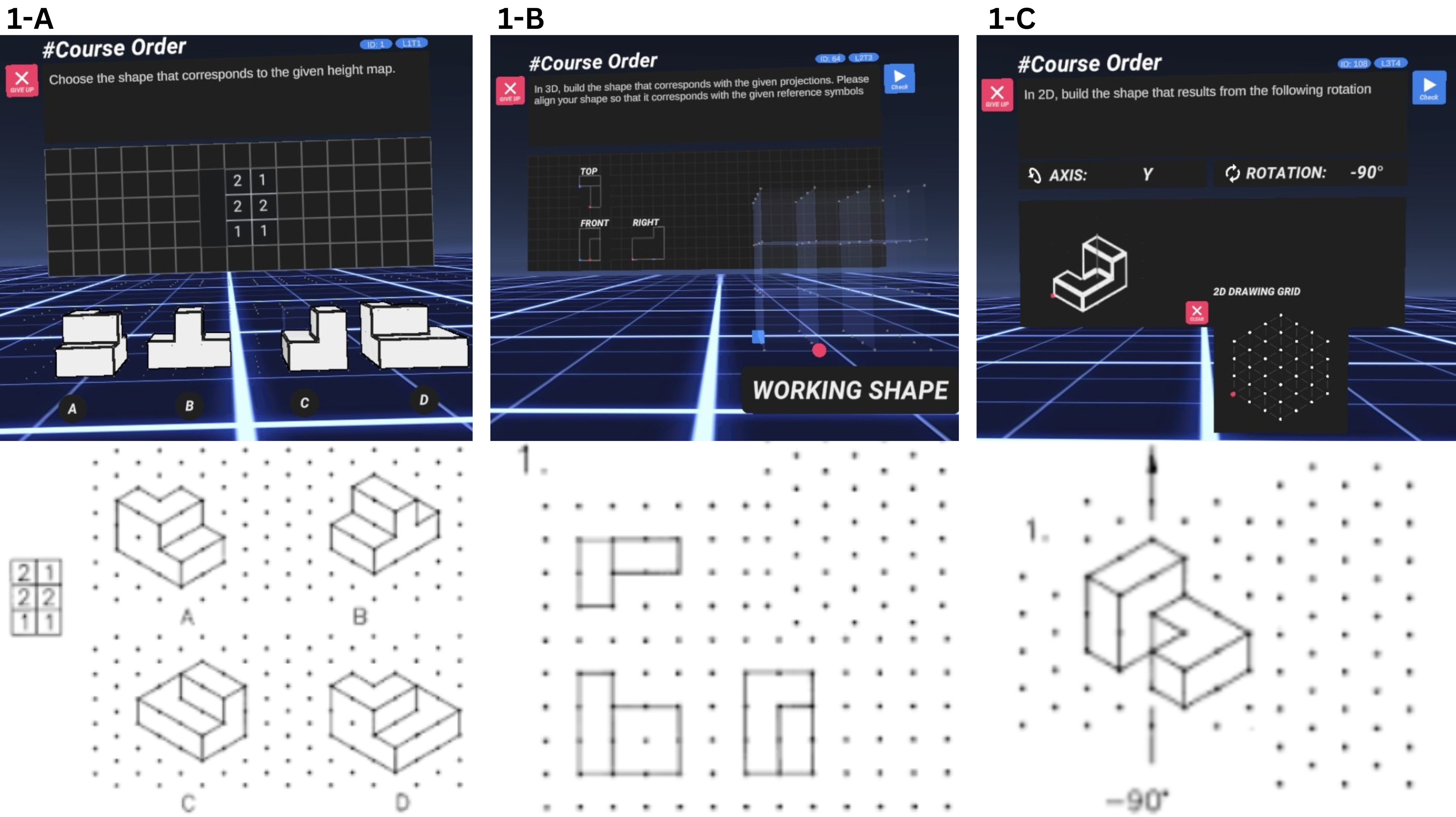}
    \caption{Examples of how problems were adapted from pencil-and-paper to VR. Figure \textbf{1-A} is a multiple choice problem from the module \textit{Isometric Sketching and Coded Plans}. Our VR instructions say ``Choose the shape that corresponds to the given height map'' where the height map refers to the height of the shape at any given point. Figure \textbf{1-B} is a drawing problem from the module \textit{Orthographic Projection}, adapted as a 3D building problem. Here users build a 3D shape by placing blocks on the grid provided. The VR instructions say ``In 3D, build the shape that corresponds to the given projections. Please align your shape so that it corresponds with the given reference symbols''. In the original course, this problem was sketched in 2D (potentially using physical snap blocks). Figure \textbf{1-C} is a 2D drawing problem from the module \textit{Rotation of Objects about a Single Axis}. The instructions say ``In 2D, build the shape that results from the following rotation''. Participants would then sketch the shape using the provided VR graph paper.}
    \label{fig:adaptation}
\end{figure*}

\subsection{\textbf{Spatial Reasoning and VR}}
\label{sec:spatial_reasoning_vr} There is evidence of a beneficial relationship between VR and spatial reasoning ability.
Simply using VR has been found to lead to increased spatial ability in certain disciplines, including architecture and astrophysics~\cite{DARWISH2023, baumgartner2022}. To our knowledge, only two prior studies have attempted to specifically train spatial reasoning skills in virtual reality. In one nine-problem study, Chang \emph{et al.} used a VR system, where users could touch physical objects in the real world that resembled the objects they were touching in virtual reality~\cite{chang2017}. 
College students were split in two intervention groups using either the VR system or a 2D computer-based system, and a control group that received no training. The VR group was the only group to have significant gains on their spatially-related post-test. In another study, Carmona \emph{et al.} guided college students through a 40-minute spatial training session using either VR, or a traditional computer screen. All students were also asked to explore the training material further at home for at least one hour in the two weeks after the training. The students in the VR group had significantly greater gains on a spatial reasoning post-test than the group that received training on a computer~\cite{Carmona2018}.
While these studies have promising results, they only investigate a single short intervention or free-form play, and did not compare against previously-validated methods of teaching spatial reasoning. By adapting the core modules of a spatial reasoning course that has been previously validated, and adding instructional feedback, we hope to improve spatial training through virtual reality in a more formal classroom setting.

\section{Application Design and Development}
In this section, we detail our design and development process for our VR application. We summarize the baseline pencil-and-paper course (Section~\ref{subsec:og_course}) and our process for adapting the content into VR (Section~\ref{sec:vr_app_info}). We also discuss our two primary additional design considerations (instructional scaffolding and interactive feedback, see Section~\ref{sec:design_elements}) and how we considered accessibility in our design (Section~\ref{sec:accessibility}). 

\label{sec:dev}
\subsection{Original Pencil and Paper Course}
\label{subsec:og_course}
To build our application, we directly adapted course material from a validated spatial training course developed by Sorby \textit{et al.}~\cite{Sorby2009}. Their course is designed for a traditional classroom and contains core modules taught each week over 10 weeks. Throughout the course, at the beginning of every class, students watch an instructional video that introduces a new concept, and then work on workbook problems on their own. Additional 2D software is provided for students to practice the concepts before attempting the workbook exercises. The workbook includes both multiple choice problems as well as sketching problems for practicing spatial visualization (see Figure~\ref{fig:adaptation}). In addition, physical snap blocks are provided to help students visualize the transition between 3D objects and their 2D representations.

\subsection{Adapting the course for VR}
\label{sec:vr_app_info}
We adapted three core modules of the course to VR: \textit{Isometric Sketching and Coded Plans}, \textit{Orthographic Projection}, and \textit{Rotation of Objects about a Single Axis}. We focused attention on these modules because consultations with the creator of the original course indicated that these three modules were the most important for improving spatial reasoning. 
For each adapted module, we re-framed the original course content into three types of questions for our VR application. For example, multiple choice questions used identical phrasing in VR but used 3D stimuli rather than 2D. Problems involving building blocks were re-framed using 3D stimuli, and problems that asked students to draw were re-framed using a VR drawing environment. 
Figure \ref{fig:adaptation} shows example problems in the baseline course workbook and how we adapted them to VR. Critically, students using VR are able to walk around the shapes and study them from different angles while working on solutions.

\subsection{VR Application: Specific Design Elements}
\label{sec:design_elements}

In our VR application, we also chose to add or enhance educationally-focused features that have been found to improve outcomes but were not formalized in the original paper-and-pencil course. The two features we focused on are real-time interactive feedback and instructional scaffolding, both of which can be facilitated by VR~\cite{Moreno2011, Belland2017, Wisniewski2020, Kleij2015}. \emph{Instructional scaffolding} is the addition of extra temporary supports that are removed over time as students progress in their learning. Scaffolding was already present in the original course, but we chose to enhance it to take advantage of the VR medium; computer-based environments have been found to be particularly useful for effective scaffolding~\cite{Belland2017}. \emph{Real-time interactive feedback} is a process through which students receive corrective information to allow them to improve their skills. The original course has limited real-time feedback built into the 2D software that students use for practice before attempting the more challenging workbook problems. Physical workbook exercises are typically graded with feedback asynchronously: it is typically too resource-intensive for instructors to provide live real-time automatic feedback for workbook problems or snap blocks for each student separately. We chose to incorporate real time interactive feedback because of the strong research supporting its effectiveness~\cite{Wisniewski2020, Kleij2015}, as well as to take advantage of the physicality and interactivity available in VR. We describe these two features in more detail below.

\textbf{Instructional scaffolding:} The pencil-and-paper course already incorporates scaffolding by ordering the workbook problems from easiest to hardest and transitioning from  multiple-choice to sketching-based questions. This allows the students to gain comfort with a concept by solving easier problems, and then improve their skills through more challenging problems. The pencil-and-paper course also provides students with snap blocks to build the shapes they see in the workbook, allowing them to first think of the problems in concrete 3D terms rather than abstract 2D terms. In our application, we not only retained the scaffolding from the original course, but also added to it by further scaffolding the transition from 3D to 2D. Facilitating this transition is the core goal of spatial reasoning training, however it can be hard to teach explicitly due to the 2D nature of traditional instructional materials~\cite{lochhead_2022}. In the baseline pencil-and-paper course, the physical snap blocks can help fill this need. However, in a classroom setting, it can be challenging for a teacher to provide individualized support and ensure that all students are using the blocks effectively. 

By teaching spatial reasoning in VR, however, we can more formally support the transition from 2D to 3D. Specifically, we start each lesson with the spatial reasoning problems in 3D rather than 2D, and slowly transition the material to 2D over the course of the VR session. In practice, we did this by first converting the majority of the original multiple-choice problems from 2D to 3D. We then formalized the transition from 3D to 2D by adding a problem type where students are asked to physically 3D build shapes in VR in response to a 2D graphic. Eventually students are asked to respond entirely in 2D. Our added scaffolding allows students to learn the course concepts and skills in 3D, and then learn how to transfer these skills to a 2D context. By having explicit 3D problems rather than optional snap blocks we are able to structure our support to ensure that it is used by all students.

\begin{figure}[tp]
    \centering
    \includegraphics[width=9cm]{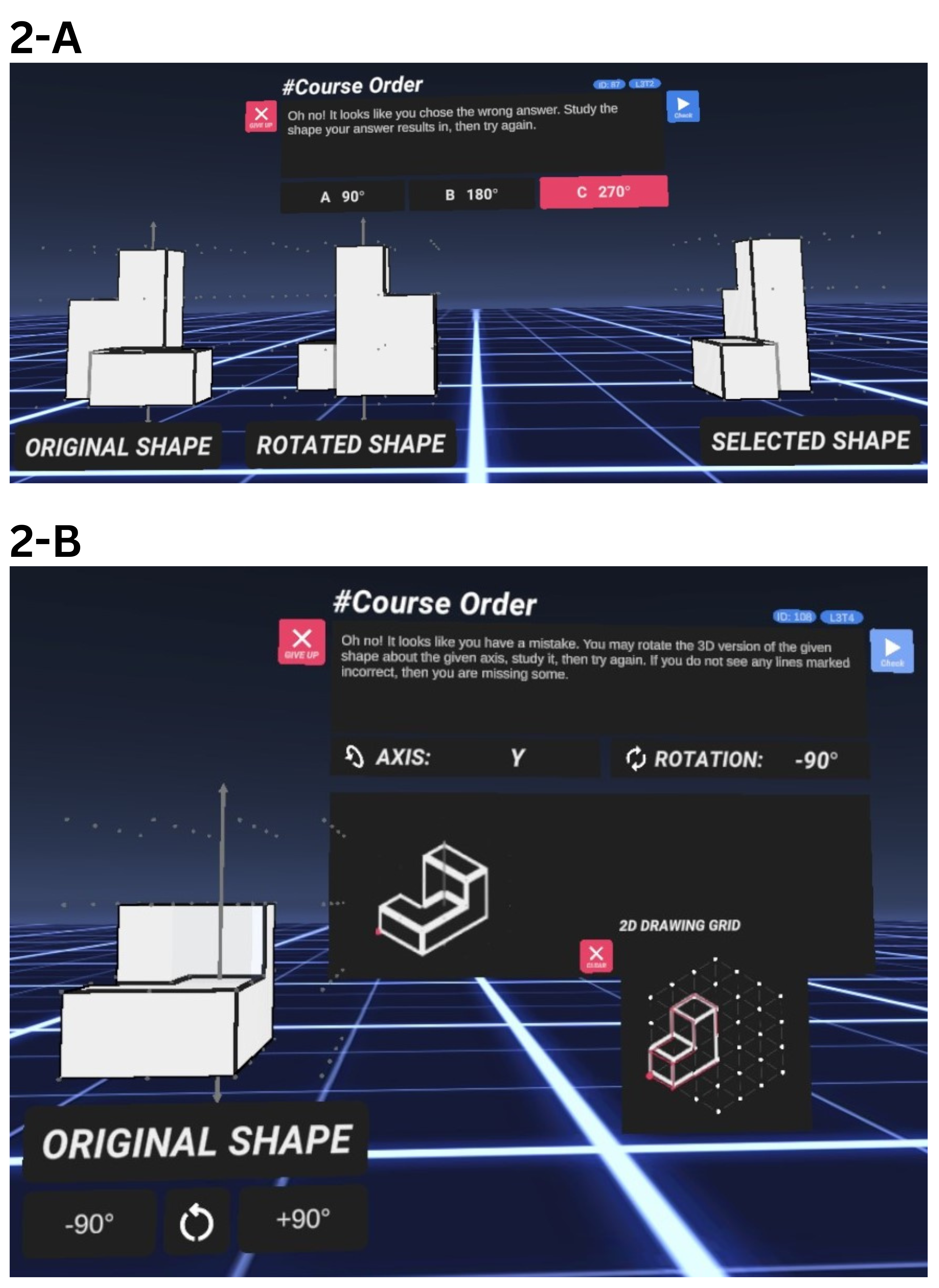}
    \caption{Feedback given for multiple choice questions (A) and 2D drawing questions (B) in the module \textit{Rotation of Objects about a Single Axis}. The text for (A) says ``Oh no! It looks like you chose the wrong answer. Study the shape your answer results in then try again''. The text for (B) says ``Oh no! It looks like you have a mistake. You may rotate the 3D version of the given shape, study it, then try again. If you do not see any lines marked incorrect, then you are missing some''. This feedback allows students to reason about how to solve the problem by seeing the solution to a similar problem.}
    \label{fig:feedback}
\end{figure}

\textbf{Interactive Real-Time Feedback:} Real-time interactive feedback is quite effective for improving student outcomes~\cite{Wisniewski2020, Kleij2015} and we propose its use in all problem types in our VR setting. The kind of feedback a student receives depends on the type of problem, and it is shown to students automatically once they submit an incorrect solution. Some of the feedback we introduced was inspired  by the feedback given in the 2D software associated with the original course. As an example, there is a type of multiple choice question that asks the student by how many degrees a shape was rotated about an axis to result into another shape. If the student answers incorrectly, they will be shown what the shape \textit{would} have looked like if it had been rotated by the amount of degrees they chose (Figure~\ref{fig:feedback}-A). This type of feedback does not give away the answer, and allows the students to understand their mistakes and learn how to reason to get to the correct solution. The ability to implement such feedback techniques is one advantage of using VR over using a traditional course setting. More details about the specific feedback of each type of problem can be found in the replication package.

\subsection{Accessibility Design Considerations}
\label{sec:accessibility}
It is well established that VR can cause cybersickness (nausea, headaches, and dizziness)~\cite{Caserman_2021}. Since experiencing these symptoms would dissuade students from using a VR application and is an accessibility concern, we aimed to address this issue in three main ways. Caserman \emph{et al.} found that joy-stick based continuous movements cause significantly more nausea than discrete movements~\cite{Caserman_2021}. Thus, we implemented \emph{teleportation}, a method of discrete movement, for those that found continuous movement nauseating. In addition, we slowed down the continuous movement. Lastly, we added a fixed background, as Park \emph{et al.} found that a moving background significantly increased the presence of cybersickness symptoms~\cite{Park_Lee_2020}. For a discussion on the effectiveness of these methods see Section~\ref{discussion}.

\subsection{Application Development}
The application was developed by working closely with the VR development team at \textit{institution removed for blind review}.
The research team gave a set of initial design documents including examples of desired feedback as well as storyboards of core application functionalities to the developers, which were then followed by monthly clarifications and progress updates. We followed an iterative development process, in which both the researchers and developers tested every iteration of the application and suggested improvements until the design goals were met. Each module was then piloted by a university student who was not part of either the research or development teams to identify any remaining bugs before deploying the application in a classroom session. During our experimental evaluation, we identified and patched one final bug that caused an issue with the the answer-checking of three problems.

\section{Experimental Design}
\label{sec:setup_design}
To evaluate our VR application we ran $n=24$ undergraduates from \textit{institution removed for blind review} through our VR spatial training course. This course consisted of three training sessions across three weeks. We measured spatial reasoning gains with a pre- and post-test design, using two previously-validated spatial tests~\cite{Yoon2011, educational1963kit}.  We also collected qualitative data on participants' VR application interactions via a post-study questionnaire. To compare our VR course to a pencil-and-paper baseline course~\cite{Sorby2009}, we compared the spatial ability gains observed in our study to those reported in an offering of the traditional version of the course at a similar institution ($n = 28$)~\cite{Endres2021}. 
In this section, we describe the structure of each VR session, give more detail on the data we collected (both qualitative and quantitative), and discuss our comparison to a baseline pencil-and-paper course.

\subsection{VR Session Structure:} Sessions occurred in a classroom equipped with 18 Oculus Rift S stations. Each student participated in three 120-minute sessions over three weeks. Before the beginning of their first session, students completed a consent and demographics form, as well as two pre-tests (see Section~\ref{sec:data_collected}). Students took the post-tests and completed a VR experience survey after the end of their third session. Multiple offerings of each module were offered each week, as determined by the research team's availability. Students generally attended their sessions on the same day and time every week, though some students chose to make up missed sessions on different days. Between two and ten participants attended any given session.

At the beginning of each session, the participants were shown instructional videos from the baseline course that described the spatial reasoning module they would be completing, accompanied by an example problem~\cite{Sorby2009}. The students then began using the VR application. 
As discussed more in Section~\ref{sec:vr_app_info}, each VR session had three types of problems adapted from the original course. Before each question type, a member of the research team demonstrated completing a sample problem. This allowed students to understand how to use the controls in the VR environment and the format of the questions. Students completed questions at their own pace without additional demonstrations. During each session, the research team emphasized and encouraged participants to take breaks and reminded participants that they were free to leave. A member of the research team walked around to answer questions, and if  participants were stuck on a problem, the researcher would assist by asking probing questions to encourage understanding the problem and to guide participants in the correct direction.

\subsection{Assessments and Data Collected}
\label{sec:data_collected}
The data collected in this study included two pre and post spatial tests, a demographic survey and a final qualitative survey. Additional quantitative data was logged by the VR application itself, while additional qualitative data was collected via observational notes from the research team during each session.

\textbf{Spatial Assessments and Demographic Survey:} We measured spatial gains using two validated assessments of spatial reasoning ability: the  \emph{Paper Folding Test} (PFT)~\cite{educational1963kit} and the \emph{Revised Purdue Spatial Visualization Test} (PSVT:R)~\cite{Yoon2011}. The demographic survey consisted of fifteen short answer and multiple choice questions.

\textbf{Post Study Qualitative Survey:} At the end of the course, each participant filled out a survey consisting of eleven short answer questions. These surveys were qualitatively analyzed to see whether students enjoyed learning in a virtual reality environment. Some of these questions included:
\begin{itemize}
    \item Have you ever used a VR head-mounted display before? If yes, do you use one often?
    \item Did the application/course make the material enjoyable?
    \item Was there something specific you liked about the application/training material?
    \item Was there something specific you didn't like about the application/training material?
\end{itemize}
The replication package contains the full list of questions asked in the survey.

\textbf{VR Application Data:} The application recorded each problem the participant completed along with the time taken. In addition, the time spent interacting with feedback was recorded for submissions in which a mistake was made. 

\subsection{Pencil-and-Paper Course Comparison}
\label{sec:pnp_comparison}
To determine whether students taught in virtual reality differed in spatial reasoning improvement from students taught in a traditional environment, we used historical data ($n=28$)~\cite{Endres2021} as well as a small subset of additional students ($n=4$). The 28 students from a previous study completed 6--10 modules of the baseline pencil-and-paper course, while the four new participants we ran completed only the corresponding three weeks of the course that we adapted to VR (but completed those weeks using pencil-and-paper materials).

\begin{table}[htbp]
%\caption{Demographic Information on VR Participants}
\begin{tabular}{lrrrr}
\rowcolor[HTML]{C9DAF8} 
Paper Id & Gender                      & \begin{tabular}[c]{@{}r@{}}Pre-Test Score \\ (X/30)\end{tabular} & \begin{tabular}[c]{@{}r@{}}Post Test Score \\ (X/30)\end{tabular} & Used VR               \\
1        & Woman                       & \cellcolor[HTML]{C8DAF9}23                                       & 25                                                                & Yes, few times        \\
2        & Woman                       & \cellcolor[HTML]{B4CDF6}27                                       & 29                                                                & Yes                   \\
3        & \cellcolor[HTML]{FFFFFF}Man & \cellcolor[HTML]{A4C2F4}30                                       & 30                                                                & Yes                   \\
4        & Woman                       & \cellcolor[HTML]{EBF2FD}16                                       & 17                                                                & No                    \\
5        & \cellcolor[HTML]{FFFFFF}Man & \cellcolor[HTML]{C3D7F8}24                                       & 27                                                                & No (\textless{}30min) \\
6        & \cellcolor[HTML]{FFFFFF}Man & \cellcolor[HTML]{F0F5FE}15                                       & 25                                                                & No                    \\
7        & \cellcolor[HTML]{FFFFFF}Man & \cellcolor[HTML]{AAC6F5}29                                       & 30                                                                & Yes (not often)       \\
8        & \cellcolor[HTML]{FFFFFF}Man & \cellcolor[HTML]{D7E4FB}20                                       & 22                                                                & No                    \\
9        & \cellcolor[HTML]{FFFFFF}Man & \cellcolor[HTML]{AAC6F5}29                                       & 29                                                                & No                    \\
10       & Woman                       & \cellcolor[HTML]{FFFFFF}12                                       & 23                                                                & N/A                   \\
11       & \cellcolor[HTML]{FFFFFF}Man & \cellcolor[HTML]{BED3F8}25                                       & 28                                                                & Yes (few times)       \\
12       & \cellcolor[HTML]{FFFFFF}Man & \cellcolor[HTML]{CDDEF9}22                                       & 30                                                                & Yes                   \\
13       & Woman                       & \cellcolor[HTML]{D7E4FB}20                                       & 21                                                                & Yes (not often)       \\
14       & \cellcolor[HTML]{FFFFFF}Man & \cellcolor[HTML]{BED3F8}25                                       & 29                                                                & Yes                   \\
15       & Woman                       & \cellcolor[HTML]{AFC9F6}28                                       & 28                                                                & No                    \\
16       & \cellcolor[HTML]{FFFFFF}Man & \cellcolor[HTML]{A4C2F4}30                                       & 29                                                                & Yes (few times)       \\
17       & \cellcolor[HTML]{FFFFFF}Man & \cellcolor[HTML]{AAC6F5}29                                       & 29                                                                & Yes (few times)       \\
18       & Man                         & \cellcolor[HTML]{F0F5FE}15                                       & 16                                                                & Yes (few times)       \\
19       & Woman                       & \cellcolor[HTML]{F0F5FE}15                                       & 23                                                                & No                    \\
20       & Man                         & \cellcolor[HTML]{AAC6F5}29                                       & 30                                                                & No                    \\
21       & Woman                       & \cellcolor[HTML]{A4C2F4}30                                       & 30                                                                & No                    \\
22       & N/A                         & \cellcolor[HTML]{C8DAF9}23                                       & 22                                                                & No (one time)         \\
23       & N/A                         & \cellcolor[HTML]{CDDEF9}22                                       & 28                                                                & Yes (one time)        \\
24       & Man                         & \cellcolor[HTML]{AFC9F6}28                                       & 26                                                                & Yes (not often)      
\end{tabular}
\caption{Demographic information for all students who completed the VR study. The Pre-Test Score column is shaded based on the incoming spatial reasoning score. The lighter shading indicates lower scores.}
\label{tab:pop_context}
\end{table}

\subsection{Participant Recruitment and Population Contextualization}
\label{sec:recruitment}
\subsubsection{Participant Recruitment}
Participants were recruited from two introductory computing classes at a large public American University. Recruitment was done via in-class presentations and posts on the course forum. Out of the 50 students that signed up to attend sessions, 40 attended at least one, and 24 completed all three. We ran four additional participants through three weeks of the baseline pencil-and-paper course. Of the VR participants, 12 completed the course in the Fall semester, and another 12 did so in the Spring semester. The pencil-and-paper participants all completed the course in the Fall semester.
\subsubsection{Population Contextualization}
\label{subsec:pop_contextualization}
To help contextualize our population, we present VR participant demographic information in Table~\ref{tab:pop_context}.
As reported in ~\ref{sec:recruitment}, all students were recruited from the same two introductory computer science courses at a large American university. Of the 22 VR participants who filled out the demographic survey and completed the study, eight were women (36\%). The average age was 18.5 years and the average programming experience was 2.1 years. Fifteen out of the twenty three students who answered the question of whether they had used a virtual reality headset had used one at least once before the beginning of the study (65\%). Our students generally had high incoming spatial ability as discussed in Section~\ref{discussion}. No one had taken a spatial course prior to this study, and one student had previously taken a spatial test. 

\section{Analysis Methodology}
\label{sec:analysis_methodology}

We now discuss our process for data cleaning, along with our statistical and qualitative analysis methods.

\subsection{Data Pre-processing}

To pre-process our data, we first removed the participants that did not  complete all three training sessions, leaving 28  participants (24 VR participants, and four supplementary pencil-and-paper participants). 

The pre and post spatial assessments, demographics survey, and qualitative responses were collected via physical paper worksheets filled out by participants. We transferred all data to a digital form: for the spatial tests, we hand graded participants' multiple choice questionnaires to determine their scores. For the demographics and qualitative data, the first author transcribed all responses into a spreadsheet.

The data recorded by the VR application required additional cleaning because some participants saved their work multiple times per session, creating a new data file each time. The first author manually went through the files to ensure that only the newest file for each participant was included in the analysis of the data. One participant who failed to save work for two out of three sessions was excluded from the analysis of the application data. The data for time spent on feedback only included information about 3D building and 2D drawing problems (no multiple-choice problems) because a bug caused the information to not be recorded for multiple-choice questions.

\subsection{Statistical Methods}
\label{subsec:stat_method}
We conducted our analysis in Python, using \texttt{Pandas}~\cite{pandas}, \texttt{Scipy}~\cite{SciPy}, and \texttt{Statsmodels}~\cite{statsmodels}. All code used to analyze the data can be found in the replication package.\footnote{Replication package submitted in supporting material}
We used parametric tests in our statistical analysis. In each case, we used QQ-plots to assess normality and ensure their use was appropriate~\cite{qqplot}. Specifically, we used a paired $t$-test between pre- and post-test scores to determine whether the interventions had an effect, as well as an independent $t$-test to compare the VR and pencil-and-paper interventions. To determine the effect size of each intervention, we computed Cohen's $d$~\cite{Cohen_1988}. To analyze the application data, we again use independent $t$-tests to investigate whether application interactions differ between students with lower and higher incoming spatial ability. We defined lower incoming spatial ability as having scored below average on either pre-test. Lastly, we correct for multiple comparisons across all statistical tests using the Benjamini-Hochberg method for controlling the false discovery rate~\cite{bh}. We consider results significant if the corrected $p$ is less than $0.05$.

\subsection{Qualitative Methods}
\label{subsec:qual_methods}
The results from the eleven question VR experience survey were formally qualitatively analyzed to assess participant experiences with the application as well as the limitations of the course. The responses from each survey were collected and initially coded by one member of the research team who identified the main themes present in the surveys. A second member of the research team coded the responses and then categorized the codes into larger themes. The hierarchical categorization was looked over by a third member of the research team. The qualitative codes are presented in Table~\ref{tab:qaul_quotes}. 

\section{Research Questions and Results}
\label{sec:Results}

%Scores table
\begin{table}
\centering
%\caption{Scores on Spatial Reasoning Tests}

\begin{tabular}{@{}lrrrrrrr@{}}
\toprule
 &
  \multicolumn{1}{r}{\begin{tabular}[r]{@{}r@{}}Pre\\     Avg.\end{tabular}} &
  \multicolumn{1}{r}{\begin{tabular}[r]{@{}r@{}}Pre\\    Stdev.\end{tabular}} &
  \multicolumn{1}{r}{\begin{tabular}[r]{@{}r@{}}Post\\     Avg.\end{tabular}} &
  \multicolumn{1}{r}{\begin{tabular}[r]{@{}r@{}}Post\\    Stdev.\end{tabular}} &
  \multicolumn{1}{r}{$p$} &
  \multicolumn{1}{r}{BH $p$} &
  \multicolumn{1}{r}{$d$} \\ \midrule
\multicolumn{8}{l}{Mental Rotation Test (Out of 30)}                 \\ \midrule
VR    & 23.58 & 5.67 & 26.08 & 4.15 & \textbf{0.0020}  & \textbf{0.007} & 0.50  \\
Paper & 17.97 & 5.15 & 20.34 & 4.54 & \textbf{0.0020}  & \textbf{0.007} & 0.49 \\ \midrule
\multicolumn{8}{l}{Paper Folding Test (Out of 20)}                   \\ \midrule
VR    & 14.75 & 3.17 & 16.54 & 3.12 & \textbf{0.0003} & \textbf{0.002} & 0.57 \\
Paper & 13.38 & 3.00    & 14.59 & 3.62 & \textbf{0.0070}  & \textbf{0.015} & 0.37 \\ 
\end{tabular}
\caption{Spatial reasoning test scores. Using paired $t$-tests and correcting for multiple comparisons with the Benjamini-Hochberg (BH) method for controlling the false discovery rate we observed significant gains on a Mental Rotation test as well as a Paper Folding test in both treatments. Both interventions had a medium effect size (Cohen's $d$) on Mental Rotation and Paper Folding.}
\label{tab:scores}
\end{table}

%Gains table
\begin{table}
\centering
%\caption{VR vs. Paper Gains on Spatial Tests}

\begin{tabular}{@{}lrrrrlr@{}}
\toprule
 &
  \multicolumn{1}{c}{\begin{tabular}[c]{@{}c@{}}VR\\ Gains\\ Avg.\end{tabular}} &
  \multicolumn{1}{c}{\begin{tabular}[c]{@{}c@{}}VR\\ Gains\\ Stdev.\end{tabular}} &
  \multicolumn{1}{c}{\begin{tabular}[c]{@{}c@{}}Paper\\ Gains\\ Avg.\end{tabular}} &
  \multicolumn{1}{c}{\begin{tabular}[c]{@{}c@{}}Paper\\ Gains\\ Stdev.\end{tabular}} &
  \multicolumn{1}{c}{$p$} &
  \multicolumn{1}{c}{BH $p$} \\ \midrule
Mental Rotation &
  2.5 &
  3.56 &
  2.38 &
  3.90 &
  0.90 &
  0.90 \\ \midrule
Paper Folding &
  1.79 &
  2.06 &
  1.22 &
  2.38 &
  0.35 &
  0.38 \\ 
\end{tabular}
\caption{VR vs. paper gains on spatial tests. Using independent $t$-tests and correcting for multiple comparisons with the Benjamini-Hochberg (BH) method for controlling the false discovery rate we did not observe a significant difference in gains between the VR course and the traditional method of instruction.}
\label{tab:gains}
\end{table}

We conducted the experiment described in Section~\ref{sec:setup_design} to address three research questions:
\begin{itemize}
    \item \textit{RQ1---Efficacy:} Can students learn spatial reasoning in virtual reality? 
    \item \textit{RQ2---Comparison:} Do students in the fewer-contact-hour VR course improve as much, or more than, those in the pencil-and-paper course?
    \item \textit{RQ3---Interaction:} How do students interact with the application? Do the interactions of students with lower incoming ability differ from those with higher incoming ability?

\end{itemize}

\subsection{\textbf{RQ1---Efficacy}}
To determine whether the virtual reality intervention was successful in increasing participants' spatial abilities we used a paired $t$-test, comparing spatial assessment pre-test scores to post-test scores. As shown in Table \ref{tab:scores}, we observed a significant main effect for both mental rotation and paper folding (Mental Rotation: $t(23)=3.4$, $p=0.002$, BH $p=0.007$ and Paper Folding: $t(23)=4.2$, $p=0.0003$, BH $p=0.002$), suggesting that the intervention was successful. Further analysis using Cohen's $d$~\cite{Cohen_1988} revealed the effect to be of medium size (Mental Rotation: $d=0.5$, Paper Folding: $d=0.57$).

\subsection{\textbf{RQ2---Comparison:}}
A comparison between our three-week VR intervention and the full ten-week pencil-and-paper baseline may illuminate VR's efficiency and effectiveness for this task.
We conducted an independent $t$-test to compare the gains (post-test score minus pre-test score) of participants in the VR intervention to those in the pencil-and-paper intervention. As seen in  Table~\ref{tab:gains}, there was no significant difference for either mental rotation or paper folding 
(Mental Rotation: $t(54)=0.12$, $p=0.90$, BH $p=0.90$ and Paper Folding: $t(54)=0.94$, $p=0.35$, BH $p=0.37$), suggesting that the 360-minute VR training is as effective as the 900-minute pencil-and-paper training. These results do not change if the four pencil-and-paper participants that only received three weeks of training are excluded from analysis. The groups had significantly different mental rotation pre-test scores but did not differ significantly in their paper-folding pre-test scores (Mental Rotation: $t(54)=3.86$, $p=0.0003$, BH $p=0.002$ and Paper Folding: $t(54)=1.66$, $p=0.1$, BH $p=0.13$). 
We discuss possible reasons for this in Section~\ref{discussion}. Critically, these results suggest that spatial reasoning can be taught in VR more efficiently than with traditional methods, making it easier for students to receive this training. 

\subsection{\textbf{RQ3---Interaction:}}
\begin{figure}[tp]
    \centering
    \includegraphics[width=9cm]{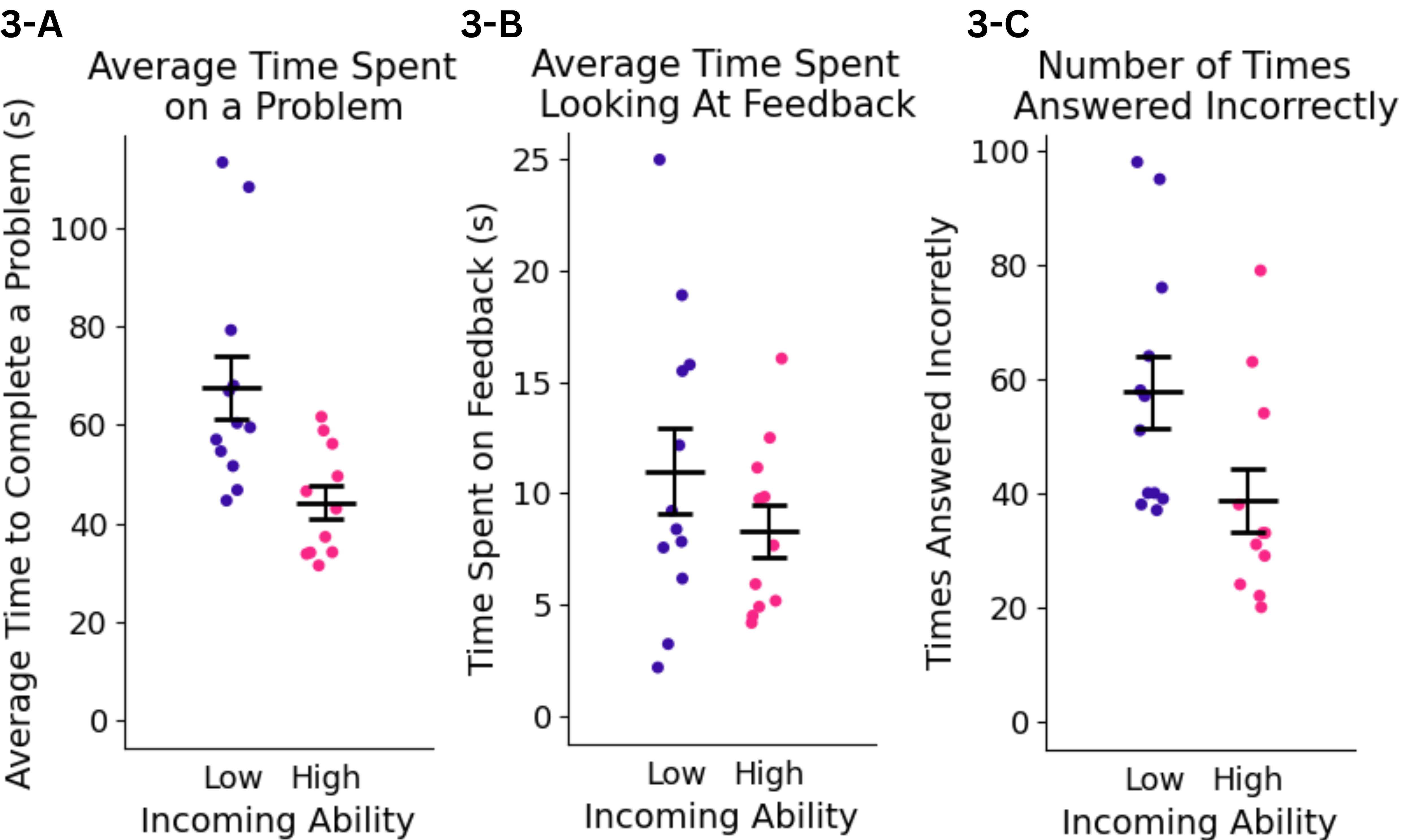}
    \caption{Interactions as a function of incoming spatial ability. The black bars indicate the mean and standard error. Using independent $t$-tests and the Benjamini-Hochberg (BH) method, we observed a significant difference in the average time spent on a problem between low and high incoming spatial ability students ($p=0.005$, BH $p=0.01$). There was no significant difference in the average time spent looking at feedback ($p=0.26$, BH $p=0.31$) and the difference in number of mistakes made was not significant after correction for multiple comparisons ($p=0.036$, BH $p=0.059$).}
    \label{fig:app_data}
\end{figure}

The VR application recorded timing, accuracy and feedback usage information. For each of these, we conducted an independent $t$-test to compare students with lower incoming spatial ability to those with higher incoming spatial ability. A student was placed in the lower incoming ability group if they scored below average on either pre-test. As seen in figure \ref{fig:app_data}, we found that students with lower incoming ability spent significantly more time solving each problem ($t(21)=3.12$, $p=0.005$, BH $p=0.01$), but did not spent significantly more time looking at feedback ($t(21)=1.15$, $p=0.26$, BH $p=0.31$). This was despite the fact that students with lower incoming ability made more mistakes, and thus had more opportunities to spend time on feedback. However, the difference in number of mistakes is not significant after correcting for multiple comparisons ($t(21)=2.23$, $p=0.036$, BH $p=0.059$). Students with lower incoming ability had significantly more gains on the mental rotation test than students with higher incoming ability ($t(21)=3.86$, $p=0.0003$, BH $p=0.002$), though that is likely due to a ceiling effect. This difference was not observed on the paper folding test ($t(21)=1.66$, $p=0.1$, BH $p=0.15$) likely because fewer participants received a perfect or nearly-perfect score on it.

We analyzed the qualitative codes as detailed in Section~\ref{subsec:qual_methods}. Based on the qualitative results from the eleven-question VR experience survey, the codes were subdivided into aspects of VR that the participants liked, VR aspects that the participants disliked, course related aspects, feedback from the program and physical discomfort.
 
\textbf{Liked VR:} We found that 20 out of 24 participants reported that they enjoyed learning in VR. Out of these 20 participants, 17 said they would take a spatial reasoning class that used VR, and 9 said they would choose VR over traditional methods. For example, one student said ``I think that the application made the material more enjoyable. I found the drawing and building blocks especially entertaining which helped me learn'' (Participant 24). Participants liked the physicality of the VR application as it allowed for easier visualization and hands-on learning (13 participants). Three participants noted that they would have liked even more visualization: ``Maybe having the ability to look at a shape from above could help visualize the shape better. Other than that, I don't have any other suggestions.'' (Participant 21). These observations suggest that students would generally enjoy, and be accepting of a VR spatial reasoning course.

\textbf{Disliked VR:} The main negatives participants noted were challenges with VR user-interface (8 participants). Four participants noted that they would not take a spatial reasoning class using VR or preferred a different method of instruction. Three participants disliked the UI, four participants found aspects of the UI to be difficult and three participants wanted changes with the UI: ``I did not like the UI for the [drawing] questions, as it was a bit tedious'' (Participant 7). This suggests that improvements can be made to VR tools to ensure that students are not hindered by difficulties unrelated to course material, such as struggling with the user-interface.

Only two participants said they would not take a VR course because of cybersickness symptoms (both cited headaches as the cause). Although this represents only $8.6\%$ of the students that completed the survey, it is vital that we explore new ways to mitigate cybersickness symptoms.

\textbf{Course-Related Aspects:} Many participants discussed other aspects of the course and learning process outside of the VR application. Seventeen participants felt that the course got easier over time. Of the 22 students that talked about session length, 18 felt that the session length was good: ``I thought it was an appropriate length'' (Participant 9). Four participants felt that the course was too long. This further supports efforts to teach spatial skills
efficiently, using fewer minutes. The time students took to complete each session is further discussed in Section~\ref{discussion}.

Six students wanted more support, and felt the material was too difficult. ``Maybe more exercises that help prepare for the 2D drawing ones. Or maybe more tools like in the final 2D drawing you could rotate the shape yourself'' (Participant 22). This observation suggests that we should continue to develop new support mechanisms such as the instructional feedback provided (cited by the student as an example of what was helpful) to further assist the students that struggle the most. 

\textbf{Physical Discomfort:} Ten students reported feeling some sort of physical discomfort. Dizziness was the highest reported (4 participants), followed by nausea (3 participants), headaches (2 participants), queasiness (1 participant), lack of physical space (1 participant) and general physical discomfort (1 participant). One student said ``The only criticism I have is that moving around with both joysticks was quite nauseating but I acknowledge that it could be limited to myself and an innate feature of VR and not a defect or flaw of the application itself.'' (Participant 14). Of the eight participants who had never used a VR headset before, four ($50\%$) reported some kind of physical discomfort as compared to the six out of the sixteen participants ($37.5\%$) who had used a VR headset at least once. Although  no formal conclusion can be drawn, this may suggest that having used VR even a few times could potentially reduce the experience of cybersickness symptoms on subsequent uses. The implications of these observations are discussed in Section~\ref{discussion}.

\textbf{Difference between lower-scoring and higher-scoring participants:} As described in Section~\ref{subsec:stat_method}, a participant was considered lower scoring if they scored below average on either spatial reasoning pre-test. There were 12 low incoming ability participants, four of whom felt that the sessions were too long ($33.3\%$) as compared to zero of the higher incoming ability participants. This is consistent with our quantitative results which showed that lower incoming ability students took significantly more time to complete each problem. This suggests that future research should explore possible ways to improve lower incoming ability students' experiences, such as breaking up each module into multiple sessions or reducing the difficulty of some problems.
    
\begin{table*}[htbp]
%\caption{Qualitative Analysis of VR Participant Feedback}
\begin{tabular}{|l|r|r|}
\hline  
\rowcolor[HTML]{C9DAF8} 
\multicolumn{1}{|c|}{\cellcolor[HTML]{C9DAF8}\textbf{Interactions}}                           & \textbf{Participant IDs}                                                                & \multicolumn{1}{c|}{\cellcolor[HTML]{C9DAF8}\textbf{Quotes}}                                                                                                                                                                                                                     \\ \hline
\rowcolor[HTML]{DAE8FC} 
\textbf{Liked VR}                                                                             &                                                                                         &                                                                                                                                                                                                                                                                                  \\ \hline
\rowcolor[HTML]{FFFFFF} 
Easy                                                                                          & all except 8, 10, 13                                                                    & ``It was relatively easy. Walking around allowed for a bit more controlled use.'' Participant 7                                                                                                                                                                                    \\ \hline
\rowcolor[HTML]{FFFFFF} 
Enjoy VR                                                                                      & all except 4, 10, 18, 20                                                                & ``Yes, it made the material more interesting and I was able to focus better'' Participant 2                                                                                                                                                                                        \\ \hline
\rowcolor[HTML]{FFFFFF} 
Think helped improve                                                                          & \begin{tabular}[c]{@{}r@{}}1, 2, 3, 4, 5, \\ 6, 7, 8, 9, 11\end{tabular}                & ``Yes, it introduced me to new methods to improve my sptial reasoning ability'' Participant 9                                                                                                                                                                                      \\ \hline
\rowcolor[HTML]{FFFFFF} 
Like UI                                                                                       & 2, 3, 21, 24                                                                            & \begin{tabular}[c]{@{}r@{}}``I specifically liked how the laser beam helped me select things more accurately. \\ I also liked how it felt like I was looking as a big screen in space.'' Participant 21\end{tabular}                                                               \\ \hline
\rowcolor[HTML]{FFFFFF} 
Liked Reluctantly                                                                             & 3, 4, 17, 18, 20                                                                        & \begin{tabular}[c]{@{}r@{}}``I'm not sure if I would take a spatial reasoning class just because, \\ but if I had to I would want to take it with VR.'' Participant 17\end{tabular}                                                                                                \\ \hline
\rowcolor[HTML]{FFFFFF} 
Physicality                                                                                   & \begin{tabular}[c]{@{}r@{}}1, 5, 6, 7, 9, 12, 15,\\ 16, 17, 18, 19, 21, 23\end{tabular} & \begin{tabular}[c]{@{}r@{}}``I liked that i was able to actually see everything in 3D and \\ can imagine that it would be significantly \\ more difficult to visualize in two dimensions on a piece of paper.'' Participant 18\end{tabular}                                        \\ \hline
\rowcolor[HTML]{DAE8FC} 
\textbf{Disliked VR}                                                                          &                                                                                         &                                                                                                                                                                                                                                                                                  \\ \hline
\rowcolor[HTML]{FFFFFF} 
Challenges with UI                                                                            & 2, 3, 5, 7, 8, 13, 18, 22                                                               & \begin{tabular}[c]{@{}r@{}}``I would suggest making the setting something more natural like maybe a physical classroom. \\ In addition it would be nice for the program to give more assistance/guidance \\ if you're struggling to solve a problem.'' Participant 18\end{tabular} \\ \hline
\rowcolor[HTML]{FFFFFF} 
Not take class                                                                                & 8, 18, 20, 21                                                                           & \begin{tabular}[c]{@{}r@{}}``I don't think I would take a class that taught spatial reasoning \\ just because I'm not sure I would need it or find the content useful.'' Participant 21\end{tabular}                                                                               \\ \hline
\rowcolor[HTML]{DAE8FC} 
\textbf{\begin{tabular}[c]{@{}l@{}}Learning Process / \\ Course-Related Aspects\end{tabular}} &                                                                                         &                                                                                                                                                                                                                                                                                  \\ \hline
\rowcolor[HTML]{FFFFFF} 
Easier Over Time                                                                              & \begin{tabular}[c]{@{}r@{}}all except 7, 10, 13,\\  15, 16, 20, 23\end{tabular}         & \begin{tabular}[c]{@{}r@{}}``Over the course of three weeks, the application became increasingly \\ more familiar and easier to use.'' Participant 14\end{tabular}                                                                                                                 \\ \hline
\rowcolor[HTML]{FFFFFF} 
Session Length                                                                                & all except 8, 10                                                                        & \begin{tabular}[c]{@{}r@{}}``The length is fine. Since it isn't much content for the given time,\\  so it was enjoyble and fun.'' Participant 7\end{tabular}                                                                                                                       \\ \hline
\rowcolor[HTML]{FFFFFF} 
Instruction                                                                                   & 4, 22                                                                                   & ``I liked the videos and being shown how to do things.'' Participant 4                                                                                                                                                                                                             \\ \hline
\rowcolor[HTML]{FFFFFF} 
Liked Course Material                                                                         & 8, 23                                                                                   & ``I enjoyed all the brain teasers.'' Participant 23                                                                                                                                                                                                                                \\ \hline
\rowcolor[HTML]{FFFFFF} 
Wanted More support                                                                           & 4, 6, 13, 18, 19, 22                                                                    & ``Maybe watch the videos before coming to sessions so we can be prepared.'' Participant 4                                                                                                                                                                                          \\ \hline
\rowcolor[HTML]{DAE8FC} 
\textbf{Physical Discomfort}                                                                  & \begin{tabular}[c]{@{}r@{}}1, 3, 6, 9, 13, 14,\\ 17, 18, 20, 21\end{tabular}            & ``The usage of VR, although fun, was headache inducing.'' Participant 20                                                                                                                                                                                                           \\ \hline

\end{tabular}
\caption{Qualitative analysis of VR participant feedback}
\label{tab:qaul_quotes}
\end{table*}

\section{Discussion}
\label{discussion}
In this section we discuss the implications of our results and possible directions for future research. Specifically, we discuss how our course could be used, as well as accessibility concerns resulting from cybersickness and lack of access.

Our finding that spatial reasoning can be taught in VR as effectively as in a traditional course but with fewer than half the contact hours has significant implications for the use of spatial interventions. A 1-credit eleven-week course like Sorby's~\cite{Sorby2009} requires a significant time commitment, and students are unlikely to elect to take it unless required by the school. A three-week training like the one presented here would be easier to incorporate into summer bridge programs or introductory engineering course modules. Summer bridge programs (SBPs) are widely employed at universities to help under-prepared students succeed, and have been shown to be effective~\cite{Bradford_Beier_Oswald_2021}. There is likely a significant intersection between the students that participate in SBPs and the students that can benefit from spatial training, so incorporating 6 hours of spatial training in SBPs could be quite beneficial for STEM. Alternatively, the relatively few contact hours required allow our training to be used as a part of introductory engineering modules employed by engineering departments to increase retention~\cite{Kiassat_Elkharboutly}.

We also observed high variance in course completion times. Some students finished each session within an hour, while others who struggled took the full two hours. This suggests that the time it takes to complete the first session could be indicative of incoming spatial ability. Gauging a student's incoming ability is important because it can determine whether the student needs to complete the course or not. Being able to gauge their incoming ability without an in-person traditional test would allow the course to be implemented in a fully asynchronous manner (i.e., students go to their university's VR lab and complete it on their own time). Alternatively, incoming spatial ability could be determined through a VR spatial reasoning test, an example of which has been recently developed~\cite{lochhead_2022}.

If such a course were to be widely used, however, it is imperative to address cybersickness, the major accessibility concern. Of the 23 students that filled out the final survey, nine indicated that they experienced some form of cybersickness ($39\%$). This is much lower than the 60--95\% reported by Caserman \textit{et al.} in a systematic review of 49 publications, indicating that our efforts to mitigate cybersickness were partially successful (see Section~\ref{sec:accessibility}). Additionally, of the nine that experienced cybersickness symptoms, two only did so when using joy-stick movement and were able to mitigate it by using teleportation instead. Another participant only got dizzy in the first session, but experienced no symptoms in the second or third, and another participant eliminated all symptoms by putting on their glasses. One participant eliminated symptoms by taking anti-nausea medication, and another only had symptoms if they spent too long in VR without breaks. Finally, one participant suffered from general motion sickness (e.g., felt sick even in car rides). These observations suggest there can be many ways to combat cybersickness. Even so, if such a course is deployed, investigating ways to reduce cybersickness symptoms is necessary to ensure an equitable learning environment. While hardware advancements are expected to improve user experience, it is evident that design considerations can have a significant effect on whether a user experiences cybersickness. For example, joy-stick based movement could be slowed down further, or eliminated completely, encouraging users to either use teleportation or move in physical space.

Another concern is that a spatial course in VR, such as the one presented here, may have its dissemination limited by access to VR technology. We do not believe this will be an issue for many universities due to the highly-increased academic interest in VR technologies~\cite{Rashid_Khattak_Ashiq_Ur}, but it remains an issue for high schools or individuals.  
Universities have already invested in creating dedicated VR labs for teaching classes about VR development as well as for research purposes
(Section~\ref{sec:background}).

Our results suggest that VR training for spatial reasoning is extremely promising. We are particularly interested in 
improvements to the course made by decreasing the number of multiple choice questions and increasing the number of 3D building questions, taking further advantage of the VR medium. Additionally, the use of augmented reality (AR) could be explored. It is likely that AR can provide most of the same benefits as VR, but alleviate some cybersickness symptoms. Ultimately, we have shown that VR can be used to teach spatial reasoning very efficiently, and we encourage future research optimizing such offerings.

\section{Threats to Validity and Limitations}
In this section we discuss some potential threats to validity as well as limitations of our study. These include experimental control and setup concerns, generalizability concerns, and a substantial drop-out rate.

One threat to the validity of our results is that the pencil-and-paper group had a significantly lower incoming spatial ability than the VR group. The VR group had unusually high average incoming ability ($79\%$ on the pre-test). One potential correlative explanation for this difference is that the VR group consisted of $66\%$ males, while the pencil-and-paper group consisted of only $16\%$ males, and males have been shown to have a higher average incoming spatial ability~\cite{Miranda2018, Segil2016}. We believe our findings that the VR course is at least as good as the traditional course are still valid, because students with lower incoming ability had significantly larger gains than those with higher incoming ability. While this is likely due to a ceiling effect, it suggests that if the incoming difference between the groups were not present there, the gains seen in the pencil-and-paper group would decrease, and those seen in the VR group would increase, which would only make the VR course better in comparison.

Another threat to the validity of our study is the drop out rate. Since our study required commitment over only three weeks, our drop-out rate of $40\%$ was larger than expected. We believe this may have occurred because the study was conducted close to the end of each semester, when students are busy with final exams. This could influence our results about the rates of cybersickness and enjoyment since those that dropped out did not complete the final survey. The researchers actively observed for signs of cybersickness and asked participants to report any symptoms throughout the study, and to our knowledge no participant that dropped out did so because of cybersickness symptoms. One participant opted to move to the pencil-and-paper group in the first five minutes of their first VR session due to cybersickness symptoms. We also cannot rule out the possibility that participants may have dropped out due to lack of enjoyment, which is a threat to our qualitative results.

Finally, it is possible that students were not engaged with the material, which could influence our results. However, we did not observe this in the VR group since most students spent time looking at the feedback provided by the application when they answered incorrectly. We mitigated this threat by looking at multiple sources of data (e.g., in-application timings, surveys, etc.) to assess participant engagement. 

\section{Conclusion}
We present the first structured VR training curriculum for spatial ability. Spatial ability can be critical for STEM success but varies with participant background. 
We adapt a validated baseline pencil-and-paper spatial ability training course to VR. We use educational scaffolding and real-time feedback in our design. 
In a statistically-significant manner, we find that our VR training improves spatial ability, and we do not find it to be any worse than the pencil-and-paper baseline. This is particularly noteworthy as our VR training requires only 360 minutes (vs. 900 for the baseline course). Qualitatively, students enjoy learning in VR, and generally indicate they would be willing to take a VR spatial reasoning course. We encourage future research on mitigating cybersickness and optimizing the design of VR spatial ability training.

\bibliographystyle{plain}
\bibliography{citations}

\end{document}